\documentclass{IEEEtran}
\usepackage{url}
\usepackage{authblk}
\usepackage{algorithm}
\usepackage{algorithmic}
\usepackage{graphicx,amsmath,amssymb,latexsym,epsfig}
\usepackage{setspace}
\usepackage{psfrag}
\usepackage{mathtools}

\newtheorem{problem}{Problem}
\newtheorem{corollary}{Corollary}

\newtheorem{lemma}{Lemma}

\newcommand{\enp} {\hfill \rule{2.2mm}{2.6mm}}

\DeclarePairedDelimiter{\ceil}{\lceil}{\rceil}
\begin{document}
\IEEEoverridecommandlockouts
\title{Minimization of Transmission Duration of Data Packets over an Energy Harvesting Fading Channel} 
\author{F. Mehmet Ozcelik}
\author{Goksel Uctu}
\author{Elif Uysal-Biyikoglu{\thanks{This work was supported by TUBITAK through grant no. 110E252 and graduate fellowships.}}}
\affil{Dept. of Electrical and Electronics Eng., METU, Ankara 06800 Turkey\\
{mehmet.ozcelik@metu.edu.tr, uctu.goksel@metu.edu.tr, elif@eee.metu.edu.tr}}
\bibliographystyle{IEEE}
\maketitle

\def\eg{{e.g.}}
\def\ie{{i.e.}}
\onehalfspacing

\begin{abstract}
The offline problem of transmission completion time minimization for an energy harvesting transmitter under fading is extended to allow packet arrivals during transmission. A method for computing an optimal power and rate allocation (i.e., an optimal offline schedule) is developed and studied.  
\end{abstract}
\section{Introduction}
\label{sec:intro}
Energy harvesting communication systems involve transmitters being powered by environmental sources such as solar, vibration, and thermal effects, either alone or as supplement to the power drawn from a grid. The ability to supply the energy storage units from environmental sources can be very useful in distributed systems such as wireless sensor networks, and M2M networks. Recent developments in ambient energy harvesting technologies have already resulted in the practical implementation of such systems~\cite{heliomote}.

Dependence on a variable energy source (as opposed to a constant supply of power) poses interesting new challenges for the transmission of information. Optimal adaptation of transmission rate has been analyzed under various problem formulations~\cite{YaU2010}-\cite{OTYUY}. In~\cite{YaU2010}, the transmission time mimimization problem on an AWGN channel of data packets arriving at arbitrary but known time instants, using energy harvests occuring again at arbitrary, known time instants was formulated and solved. In~\cite{MAAEUHE2010} and~\cite{YYOOSO2010}, the formulation was extended to a broadcast channel with a static data pool. This was extended to cover the case of new data arriving during transmission in~\cite{FMOHEEU2011ISIT,FMOHEEU2011}. In~\cite{OO_SU_2010},~\cite{RR_VS_PV_2011} bounds on the capacity of an energy harvesting AWGN  channel were obtained. The results were extended to a fading channel in ~\cite{RSV11}. The solution of the transmission completion time minimization problem on a fading channel with a static data pool and known harvest times and channel states was reported in ~\cite{OTYUY}. 

This paper extends the formulation of~\cite{OTYUY} by relaxing the static data pool assumption, and its main contribution is to  develop an offline solution for the time minimizing {packet scheduling problem} with a rechargeable transmitter under fading conditions. The solution needs to adjust its transmission power and rate over the course of transmission with respect to  packet arrivals, as well as channel state and energy harvests. This will sometimes correspond to lowering rate (therefore the energy per bit), to work energy efficiently and prevent premature data queue idleness; and at other times, increasing the rate to take advantage of a good channel state, especially when energy is abundant. The problem statement is made precise in the next section.
\vspace{-0.1 in}
\section{System Model}
\label{sec:sm}
\begin{figure}
\begin{center}
	\includegraphics[scale=0.35]{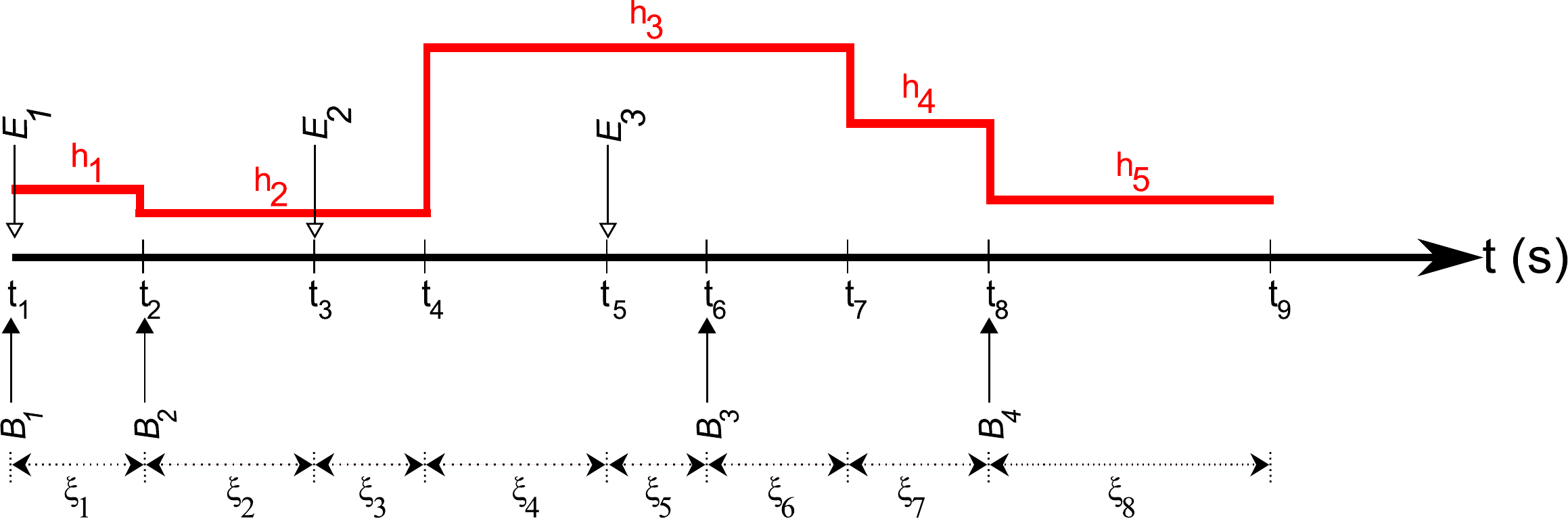}
\end{center}
\caption{An example sequence of events: $t_i$, $i\geq 1$ are event times (energy harvests marked as $E_i$, channel state changes, $h_i$, or data arrivals $B_i$). The $\xi_i$ denote inter-event epoch durations.}
\label{fig:sm}
\end{figure}

Consider point-to-point communication over a fading channel, where transmission is supplied by the harvested energy, arriving at arbitrary instants. Following the \emph{offline} formulation in~\cite{OTYUY}, we assume that the transmitter has knowledge of the energy harvests as well as channel states before transmission starts. In contrast to~\cite{OTYUY}, data packets are allowed to arrive at arbitrary (known) times during the course of transmission. The harvested energy is stored in an (ideal) battery and immediately becomes available for use by the transmitter. Data packets are stored in a data buffer (of infinite capacity.) An example sequence of energy and packet arrivals, as well as channel gain changes is illustrated in Fig.~\ref{fig:sm}. Starting from time $t_1=0$, the amounts of energy and data have become available by time $t$ are denoted by $E(t)$ and $B(t)$, respectively.  Any arrival of energy or data or a change in the channel state is called an {\emph{event}}. The duration between any two consequent events is called an epoch. The length of $i^{th}$ epoch is $\xi_i=t_{i+1}-t_{i}$. Given an average power constraint $p_i$ and channel gain level $\sqrt{h_i}$ during the $i^{th}$ epoch, we assume rate level of $r_i=\frac{1}{2}log_{2}(1+h_ip_i)$ is achievable for a certain tolerable error probability. Equivalently the power level used to transmit a codeword at rate $r_i$ is given by: $g(r_i)=\frac{2^{2r_i}-1}{h_i}$.

\vspace{-0.2 in}
We consider packets arriving in a certain time window of size $W<\infty$. The problem is to find an allocation of power and rate across time that minimizes the total duration of transmission for all of these packets.  An optimal policy should respect causality constraints (at any time, only the resources available up to that point can be used). It immediately follows from the concavity of the rate function that rate (and power) should not change within an epoch. So, the search for an optimal schedule can be limited to schedules that keep a constant power level and rate within each epoch. Hence, the optimization problem can be written in terms of rates assigned to epochs. Note that in the problem formulation below, the solution space is further limited (w.l.o.g.) to schedules spanning no more than some $k^{\rm up}$ epochs. The value of $k^{\rm up}$ can be set as the number of epochs used by any arbitrary feasible schedule.
\begin{problem}
\label{pr:FadeSchedulingTmin}   
\noindent{\bf Transmission Time Minimization of Packets on an Energy Harvesting Fading Channel:}
\small \begin{align}
\noindent \mbox{Minimize:  } &T=T(\{ r_i\}_{1 \leq i \leq k^{\rm up}})\nonumber\\
\noindent \mbox{subject to: }&r_i\geq 0  \nonumber\\
&\sum_{i=1}^{k} g(r_i)\xi_{i} \leq E(t_k), \;\; \sum_{i=1}^{k} r_i \xi_i  \leq B(t_k)\label{eq:BConst}\\
&\sum_{i=1}^{k^{*}} r_i \xi_i+ r_{k^{*}+1} (T-\sum_{i=1}^{k^{*}}\xi_{i}) = B(T)\label{eq:DConst}\\
&for \mbox{ }k=1,2,...,k^{*}=\max\{i:\sum_{j=1}^{i} \xi_j \leq T\}\nonumber
\end{align}\normalsize
\end{problem}
\vspace{-0.1 in}
In Pr.~\ref{pr:FadeSchedulingTmin}, $k^*$ denotes the last epoch used in an optimal schedule.  \eqref{eq:BConst} state the (energy and data) causality constraints, while~\eqref{eq:DConst} ensures transmission completion of all data. Following~\cite{HknThs}, we will exhibit the equivalence of Pr.~\ref{pr:FadeSchedulingTmin} to a convex problem, namely Problem ~\ref{pr:FadeSchedulingDmax}. 

\vspace{-0.15 in}
\section{An Equivalent Problem}
Problem~\ref{pr:FadeSchedulingDmax} aims to find a schedule which minimizes total energy consumption to transmit a given sequence of packets within a given deadline constraint $T$, using the energy harvested during this time. 
\begin{problem}
\label{pr:FadeSchedulingDmax}
\noindent{\bf Energy Consumption Minimization of Packets on an Energy Harvesting Fading Channel:}
\small \begin{align}
\noindent \mbox{Minimize:  }&E(T)= \sum_{i=1}^{k^{*}} g(r_i)\xi_i + g(r_{k^{*}+1})(T-\sum_{i=1}^{k^{*}}\xi_i) \nonumber\\
\noindent \mbox{subject to: }&r_i\geq 0\label{eq:rConst2}\\
&\sum_{i=1}^{k} g(r_i)\xi_{i} \leq E(t_k), \;\; \sum_{i=1}^{k} r_i \xi_i  \leq B(t_k)\label{eq:BConst2}\\
&\sum_{i=1}^{k^{*}} r_i \xi_i+ r_{k^{*}+1} (T-\sum_{i=1}^{k^{*}}\xi_{i}) = B(T)\label{eq:ECnsmd}\\
&for \mbox{ }k=1,2,...,k^{*}=\max\{i:\sum_{j=1}^{i} \xi_j \leq T\}\nonumber
\end{align}
\end{problem}\normalsize

\begin{lemma}
\label{lmm:ConvexProblem}
Problem~\ref{pr:FadeSchedulingDmax} is a convex optimization problem.
\end{lemma}
\noindent{\emph{Proof.}} Firstly, note that $g(r_i)$ is a strictly convex, monotonically increasing function. Furthermore, the constraint set of the problem is defined by non-negative weighted sums of either $g(r_i)$'s and $r_i$'s, each constraint being either convex or linear, respectively. It easily follows (please see~\cite{FthThs} for details) that the set of feasible allocations form a convex region. Finally, as the objective function of the minimization problem is also a non-negative weighted sum of increasing convex functions, we conclude that  Pr.~\ref{pr:FadeSchedulingDmax} is convex~\cite{Boyd}. 

\begin{lemma}
\label{lmm:Equivalence} Suppose $T$ is the minimum completion time (obtained by solving Pr.~\ref{pr:FadeSchedulingTmin}) for the sequence of packets arriving by time $W$, $0<W\leq T$. Then, for this sequence of events, any solution of Pr.~\ref{pr:FadeSchedulingDmax} with deadline constraint specified as $T$ provides a solution to Pr.~\ref{pr:FadeSchedulingTmin}.
\end{lemma}
\noindent{\emph{Proof.}} Let schedules $S^{1}$ and $S^{2}$ be optimal solutions of Pr.~\ref{pr:FadeSchedulingTmin}, and Pr.~\ref{pr:FadeSchedulingDmax}, defined with the deadline $T$, respectively. The energy consumption of both schedules must be the same since the opposite claim would contradict the optimality of the schedules: $S^{2}$ has used no more energy by $T$ than $S^{1}$, by definition. Suppose it used less energy, than this means that $S^{1}$ has used some extra energy to transmit the same packets as $S^{2}$. But then, it could use this extra energy in the last epoch to reduce the completion time by a nonzero amount, which would contradict optimality. Hence, we have 2 schedules completing the transmission of the same amount of data at the same time by consuming same amount of energy. Thus, $S^{1}$ and $S^{2}$ are both solutions to problems~\ref{pr:FadeSchedulingTmin} and~\ref{pr:FadeSchedulingDmax}. But by convexity, Pr.~\ref{pr:FadeSchedulingDmax} has a unique solution, hence $S^{1}$ and $S^{2}$ must be the same.\enp\\

\vspace{-0.4 in}
\subsection{Solution of Problem 2}  

The \emph{sequential unconstrained minimization technique (SUMT)} is a convenient method~\cite{luenberger,appendix} for iteratively converging to the optimal of a constrained problem by solving a sequence of unconstrained optimization problems. The unconstrained problems are formed by adding to the objective of the original problem penalty terms corresponding to constraint violations. Correspondingly in our case, we obtain Pr.~\ref{pr:Unconstrained_Minimization} as follows:

\begin{problem}
\label{pr:Unconstrained_Minimization}
\noindent{\bf Unconstrained Minimization Problem:}
\small\begin{align}
\noindent \mbox{Minimize:  } F(\textbf{r})&=\left(\sum_{i=1}^{k^*}g(r_{i})\xi_{i} + g(r_{k^{*}+1})(T - \sum_{i=1}^{k^{*}}\xi_i)\right) + \mu P(\textbf{r}),\nonumber
\end{align}
\end{problem}
\vspace{-0.2 in}
\setlength\arraycolsep{0.1 em}
\small\begin{eqnarray}
\mbox{where, }P(\textbf{r}) &=& \sum_{i=1}^{k^*+1}(\max(0,-r_{i}))^2 \label{eqn:SUMT_Nonnegativity}\\
&+& \sum_{k=1}^{k^*+1}\left(\max(0, \sum_{i=1}^{k}g(r_{i})\xi_i - E(t_k))\right)^2\label{eqn:SUMT_EnergyCausality}\\
&+& \sum_{k=1}^{k^*+1}\left(\max(0, \sum_{i=1}^{k}r_{i}\xi_i - B(t_k))\right)^2\label{eqn:SUMT_DataCausality}\\
&+& \left(\sum_{k=1}^{k^*}r_{i}\xi_i + r_{(k^*+1)}(T - \sum_{i=1}^{k^{*}}\xi_i) - B(T)\right)^2\label{eqn:SUMT_Ecnsmd}
\end{eqnarray}
\normalsize

Due to constraints~\eqref{eq:rConst2},\eqref{eq:BConst2} and~\eqref{eq:ECnsmd}, penalty terms~\eqref{eqn:SUMT_Nonnegativity},~\eqref{eqn:SUMT_EnergyCausality},~\eqref{eqn:SUMT_DataCausality} and~\eqref{eqn:SUMT_Ecnsmd} have been added to the objective function. Starting from a point in the exterior of the feasible region for an initial value of the penalty coefficient $\mu=\mu_0$, we reach the next point by solving the corresponding unconstrained minimization problem. At each SUMT iteration, initial point is moved to the previously computed result. By iterating the penalty coefficient such that after iteration $n$, $\mu^n=\eta\mu^{n-1}$ for some growth parameter $\eta\geq 1$, we solve a sequence of unconstrained problems with monotonically increasing values of the penalty coefficient. Intuitively, this drives the points toward the feasible region. It is proved in ~\cite{luenberger} that in the case of a convex objective and penalty terms as defined above, the algorithm converges to the optimum of the original constrained problem as $\mu$ goes to infinity. In practice, the iterations are stopped when an arbitrary stopping criterion $1/\mu\leq\epsilon_S$ is satisfied.

In our problem, SUMT is initialized with an infeasible allocation (i.e., at a point in the exterior of the constraint region), specifically, transmitting all data at constant rate within the given deadline $T$, disregarding causality constraints. To ensure fast convergence (see~\cite{luenberger,appendix}) $\mu$ is initialized such that the values of the objective and penalty terms are commensurate, and the penalty terms corresponding to each constraint are scaled such that no constraint dominates. At each iteration of SUMT, the corresponding unconstrained problem is solved by {\emph{Newton's method}}. It is quite standard to apply Newton's method in the inner iterations of SUMT.  After the $l^{th}$ Newton step in an inner iteration, rate allocation vector is updated as: $\textbf{r}^{l+1} = \textbf{r}^{l} - [[\nabla^2F(\textbf{r}^l)]^{-1} \nabla F(\textbf{r}^l)$. The \emph{Newton decrement}, $\lambda(\textbf{r}^l) = \left(\nabla F(\textbf{r}^l)^T [\textbf{H} F(\textbf{r}^l)]^{-1} \nabla F(\textbf{r}^l)\right)^{\frac{1}{2}}$, becoming smaller than a predefined accuracy parameter $\epsilon_N$ is the stopping criterion for each inner iteration. By reducing $\epsilon_N$, the inner optimizations can be made arbitrarily accurate~\cite{Boyd}. The convergence rate of these iterations will be discussed in the following sections.
\vspace{-0.2 in}
\subsection{Solution of Problem~\ref{pr:FadeSchedulingTmin}}  
From Lemma 2, using the optimal value of completion time, $T^{\rm opt}$, as a parameter in Pr.2 would give us an optimal schedule for Pr.1. Of course, $T^{\rm opt}$ is not known before solving Pr.1. The method we will use is to iteratively approach $T^{\rm opt}$ by solving Pr. 2 for different values of $T$ and checking the resulting amount of energy consumption. The bisection method will be used to monotonically narrow down the interval in which the optimal completion time $T^{\rm{opt}}$ of Pr.~\ref{pr:FadeSchedulingTmin} must lie in. Since $E(T)$ is a monotonically decreasing and continuous function of $T$~\cite{HknThs}, any feasible value of $T$ provides an upper bound on $T^{\rm{opt}}$. In search of upper and lower bounds, $T$ is initialized as the end of last data arrival epoch, and SUMT is run as detailed in Section A. If the resulting optimal energy that SUMT returns is too high, it means that transmission cannot be completed within this deadline, hence the current value of $T$ provides a lower bound. $T$ is then is extended by the next epoch length. This procedure is repeated until the total consumed energy returned by SUMT goes below the energy harvested by $T$. That value of $T$ provides an upper bound.  The next deadline is chosen as the average of the upper and lower bounds, and SUMT is run again. If the deadline is feasible, it becomes the new upper bound, if not, it becomes the new lower bound, and so on. The iterations are stopped when the difference between the upper and lower bounds goes below $\epsilon_b$, which, provided that the inner optimizations of SUMT are also done with sufficient accuracy, sandwiches $T^{\rm{opt}}$ in an interval of size $\epsilon_b$.
\vspace{-0.05 in}
\section{Computational Complexity}  
\label{sec:comp}

The computational complexity is largely imposed by the stopping criteria of Newton, SUMT and bisection iterations. To compute the overall complexity of proposed scheme, let us first consider the number of bisections. When bisection iterations begin, the difference between upper and lower bounds on completion time becomes the last epoch length of the most current schedule returned by SUMT. In each iteration this interval is halved, so at most $\ceil*{log_2(\xi_{k^{\star}+1}/\epsilon_b)}$ bisections are to be performed. For each bisection, SUMT makes  $\ceil*{log(\frac{1}{\mu\epsilon_S})/log(\eta)}$ iterations to converge with a desired accuracy of $\epsilon_S$~\cite{Boyd}. The number of Newton steps to achieve an accuracy of $\epsilon_N$ in the inner Newton iterations per each iteration of SUMT is upper bounded by $\frac{F(\textbf{r}^0)-F^{\star}}{\gamma}+log_2 log_2(1/\epsilon_N)$, where $\gamma$ is the minimum decrement amount of $F$ and $F^{\star}$ is the value at the optimal point~\cite{Boyd}. This bound follows from the different nature of convergence of Newton's iterations for different operating points. It has been shown in~\cite{Boyd} that, once the operation point gets sufficiently close to the optimum, convergence rate is quadratic, while it is approximately linear until then. Finally, the computational requirements imposed by each Newton step, due to the construction and the inversion of a $2k \times 2k$ Hessian (where $k$ is the number of epochs in the problem), is polynomial (with complexity $O(k^3)$ or as low as $O(k^2)$ with ultimately efficient implementation.) 
\vspace{-0.05 in}
\section{Numerical Results} 
\begin{figure}
\begin{center}
\includegraphics[scale=1.15]{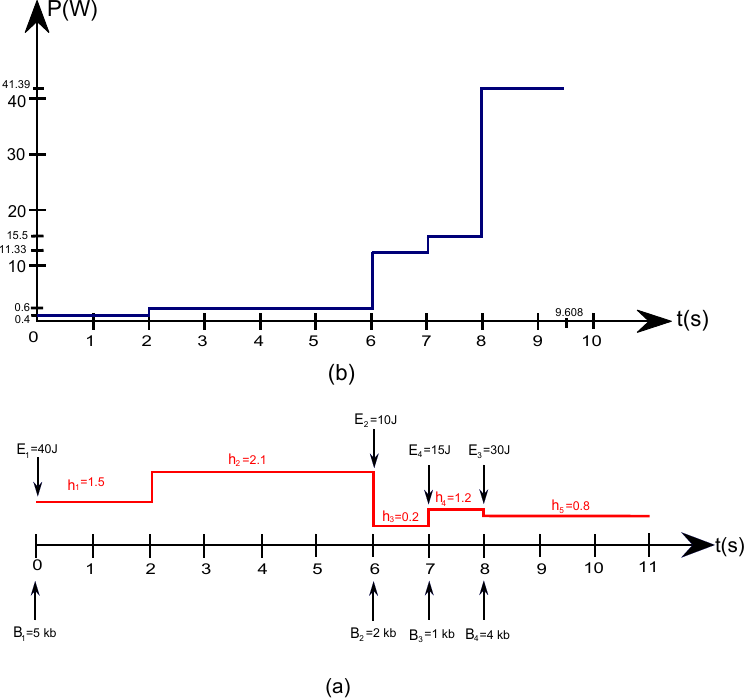}
\end{center}
\caption{(a) An example event sequence. The SNR in the $i^{th}$ epoch is $h_i$, the bandwidth is $W=1$ KHz, energy harvest amounts and arriving data  are marked as $E_i$ and $B_i$, respectively. (b) Final schedule returned by completion time minimization algorithm.}
\label{fig:NumEx}
\end{figure}

As an example, we consider the event sequence depicted in Fig.~\ref{fig:NumEx}. The final schedule returned by the proposed algorithm is also shown in the figure. The penalty parameter $\mu$ is initialized as $1$, the growth parameter $\eta$ is set to $2$.  The threshold values for Newton's method, SUMT and bisection are $10^{-8}$, $10^{-10}$ and $10^{-3}$, respectively. The algorithm repeats $34$ SUMT iterations, within which at most 6 Newton's steps are repeated, for each of $12$ bisection repetitions and terminates within $95$ seconds in MATLAB running on a MacBook Pro rev. 8.1. When the Newton's and bisection thresholds are raised to $10^{-3}$ and $10^{-2}$, respectively, the run time reduces to $15$ seconds. We believe that optimizing the code over an efficient programming platform can reduce this time significantly.
\vspace{-0.05 in}
\section{Conclusion}  
\label{sec:conc}

A method for solving the offline minimum completion time packet scheduling problem on an energy harvesting fading channel has been developed and demonstrated. The key to the method is exhibiting equivalence to an energy minimization problem which is a convex program. In certain realistic scenarios, the harvest profile and data arrivals may be known in advance. In that case, the offline solution would apply for a static channel. On a fading channel with an ergodic channel state process, an online algorithm such as waterfilling could run on top of the offline adaptation. When the data and/or harvest arrivals are also unknown, the offline solution here may be combined with a prediction or learning scheme or a simple look-ahead policy.
\vspace{-0.05 in}
\singlespacing

\end{document}